\RequirePackage{amsmath}
\documentclass[twocolumn,twocolappendix]{aas}
\usepackage[utf8]{inputenc}
\usepackage[varg]{txfonts}
\usepackage{gensymb}
\usepackage{nameref}
\usepackage{multirow}
\usepackage{listings}
\usepackage{lipsum}
\usepackage{lmodern}
\usepackage{breqn}
\usepackage{outlines}
\hypersetup{linkcolor=blue,citecolor=blue,filecolor=blue,urlcolor=blue}

\newcommand{\giv}{\text{\textbar}}

\begin{document}

\title{\large Reconciling the Systemic Kicks of Observed Millisecond Pulsars,\\\vspace{1mm}Spider Pulsars, and Low-mass X-ray Binaries}
\shorttitle{Reconciling the Systemic Kicks of Observed MSPs, Spider Pulsars, and LMXBs}

\author[0000-0002-0492-4089]{\normalsize Paul Disberg}
\affiliation{School of Physics and Astronomy, Monash University, Clayton, Victoria 3800, Australia}
\affiliation{The ARC Centre of Excellence for Gravitational Wave Discovery---OzGrav, Australia}
\email[show]{\href{mailto:paul.disberg@monash.edu}{paul.disberg@monash.edu}}

\author[0000-0003-2506-6041]{\normalsize Arash Bahramian}
\affiliation{International Centre for Radio Astronomy Research, Curtin University, GPO Box U1987, Perth, WA 6845, Australia}
\email{}

\author[0000-0002-6134-8946]{\normalsize Ilya Mandel}
\affiliation{School of Physics and Astronomy, Monash University, Clayton, Victoria 3800, Australia}
\affiliation{The ARC Centre of Excellence for Gravitational Wave Discovery---OzGrav, Australia}
\email{ilya.mandel@monash.edu}

\shortauthors{Disberg, Bahramian, and Mandel}

\received{January 18, 2025}
\revised{-}
\accepted{-}

\begin{abstract}
\noindent Millisecond pulsars (MSPs) have been proposed as evolutionary products of low-mass X-ray binaries (LMXBs) through a stage in which they are spider pulsars (i.e., redbacks and black widows). However, recent work has found that the systemic kicks of observed MSPs are significantly lower than the kicks of LMXBs and spiders, which appears to be in tension with this evolutionary model. We argue that this tension can be relieved, at least to some degree, by considering the fact that the observed MSPs are located at relatively short distances, whereas spider pulsars are located at greater distances and LMXBs are situated even further away. We model the distance-dependent kinematic bias for dynamically old objects, which favors observing objects that have received low kicks at short distances and correct the observed systemic kicks for this bias. We find that this kinematic bias can be big enough to close the gap between the MSP and LMXB kicks, although the spider pulsars appear to come from a slightly different systemic kick distribution, but this difference is not necessarily physical. All corrected systemic kick distributions are consistent with predictions from binary population synthesis for progenitor systems with a post-supernova orbital period of $P_{\text{orb}}\leq10\,$d and a companion mass of $M_{c}\leq1\,M_{\odot}$, where the natal kicks are calibrated to the velocities of young isolated pulsars. We conclude that the difference in observed systemic kicks is not necessarily in tension with a common origin for MSPs, spider pulsars, and LMXBs.
\end{abstract}

\section{Introduction}
\noindent Millisecond Pulsars (MSPs) are old pulsars \citep{Toscano_1999} that spin rapidly. These objects are thought to be born as regular pulsars that later spin down and cease radio emission, after which they were spun up again by accretion from a companion star, recycling the neutron star (NS) into an MSP \citep{Alpar_1982}. 

One subcategory of MSPs is the class of ‘spider’ pulsars, binary systems consisting of an MSP and a very low-mass companion \citep{Chen_2013,Roberts_2013}. In these systems, the companion stars are ablated by the MSP \citep{Kluzniak_1988}, meaning they experience irradiation-driven mass loss. The spider pulsars have binary orbital periods of $P_{\text{orb}}\lesssim1\,$d and are categorized into two classes: `redback' pulsars, with companions of $M_{c}=0.1{-}0.4\,M_{\odot}$, and `black widow' pulsars, with $M_{c}\ll0.1\,M_{\odot}$ \citep{Chen_2013}. There is currently no consensus about the evolutionary connection between redbacks and black widows; while some argue that redbacks evolve into black widows, others find that they share a common origin after which they diverge (see, e.g., \citealt{King_2003}; \citealt{DeVito_2020}; \citealt{Ginzburg_2020,Ginzburg_2021}; or the discussion in \citealt{ODoherty_2023}).

Meanwhile, low-mass X-ray binaries (LMXBs) that contain an NS and a low-mass companion ($M_c\leq1\,M_{\odot}$) are observed through X-ray emission caused by mass transfer onto the NS \citep{VandenHeuvel_1975}. This mass transfer is expected to facilitate pulsar recycling \citep{Bhattacharya_1991}. LMXBs are relatively old \citep{Cowley_1987} and have tight orbits ($P_{\text{orb}}\lesssim10\,$d).

A natural explanation for observed isolated MSPs is that LMXBs can evolve into spider pulsars, where the ablation is strong enough to eventually evaporate the companion completely. In this model, MSPs, spiders, and LMXBs have a common progenitor system in which an NS forms without unbinding the binary, imparting a kick velocity on the system \citep[e.g.,][]{Tauris_1996}. Isolated NSs receive typical natal kick velocities of several hundred km\,s$^{-1}$ on their formation \citep{Verbunt_2017,Igoshev_2021,Disberg_2025a,Disberg_2025b} as a consequence of asymmetry in the supernova explosion \citep[e.g.,][]{Janka_1994,Burrows_1996,Müller_2019,Burrows_2024}. Meanwhile, binary progenitor systems experience a total systemic kick which is a combination of the natal kick and a Blaauw kick due to the recoil after mass loss in a supernova \citep{Blaauw_1961,Brandt_1995}. If MSPs, spider pulsars, and LMXBs are indeed evolved from the same progenitor systems, one should expect these objects to have received similar systemic kicks. We note that, although most MSPs are thought to form through this channel, some of them are formed with high mass companions, such as the double-pulsar system J0737--3039 \citep{Burgay_2003}.

\citet{ODoherty_2023} estimated the systemic kicks of observed MSPs, black widows, redbacks, and LMXBs in the Galactic field. They assembled a catalogue of 145 systems that have measured distances, proper motions, and possibly radial (i.e., line-of-sight) velocity. For systems without known radial velocity, they assumed the peculiar velocity to be distributed isotropically. Then, they employed the method of \citet{Atri_2019} to estimate the systemic kicks by integrating the orbits of the systems back in time through the Galactic potential, where the posterior on the kick distribution is given by the peculiar velocities at the disc crossings (assuming these systems were born in the Galactic thin disc). While \citet{ODoherty_2023} find similar systemic kicks for their samples of LMXBs, redbacks, and black widows, the systemic kicks for their sample of MSPs are significantly lower.

Their method is able to constrain the kicks of these relatively old objects, despite the fact that their present-day velocity is no longer representative of their initial kick \citep{Disberg_2024a,Disberg_2024b}. It is assumed that the present-day systemic velocity is only a result of the natal kick and subsequent interaction with the Galactic potential. This means that the pre-supernova peculiar velocity is assumed to be zero, while the peculiar velocities of young massive stars in the thin disc are on the order of ${\sim}\,10\,$km\,s$^{-1}$. Moreover, it is assumed that there are no hypothetical ``rocket" kicks \citep{Harrison_1975} that affect the systemic velocities after the supernova. Although it may be possible that rocket kicks reach magnitudes similar to natal kicks \citep{Hirai_2024}, only low-magnitude rocket kicks are proposed as explanation for, e.g., wide neutron star binaries observed by Gaia \citep[${\lesssim}\,30\,$km\,s$^{-1}$,][]{Baibhav_2025}. Pre-supernova velocities or hypothetical rocket kicks are therefore unlikely to affect the kick estimates significantly.

The results of \citet{ODoherty_2023} raise the question of whether the lower kick velocities of the observed MSPs suggest that the model where LMXBs evolve into spiders and ultimately into MSPs does not provide the full picture. We argue, however, that the lower MSP systemic kicks can be explained by the fact that they are observed at significantly shorter distances than the other objects, despite receiving the same systemic kicks (Section \ref{sec2}). To illustrate this point, we employ the simulation of \citet{Disberg_2025a}, which contains Galactic trajectories of kicked objects, to analyze the relationship between kick magnitude and distance to the Sun (Section \ref{sec3}). This allows for comparison between the observed systemic kicks and a prediction based on the distances to these systems and an assumed intrinsic systemic kick distribution (Section \ref{sec4}). Through this comparison, we can find a systemic kick distribution that can explain the observations (Section \ref{sec5}), in an attempt to reconcile the systemic kicks of MSPs, spider pulsars, and LMXBs (Section \ref{sec6}). 

\section{Observation}
\label{sec2}
\noindent The sample of \citet{ODoherty_2023} consists of 95 MSPs \citep{Manchester_2005}, 17 black widows \citep{Hui_2019}, 14 redbacks \citep{Strader_2019}, and 19 (NS) LMXBs \citep{Arnason_2021}. They only include systems with precise distance and proper motion measurements, which are available for only a small number of systems \citep[cf.\ the LMXB catalogue of][]{Fortin_2024}. Since their method assumed formation in the Galactic thin disc, they exclude systems that are associated with globular clusters, where short-period systems can be formed dynamically \citep{Pooley_2003}. The distances to all MSPs and black widows are estimated through dispersion measure (DM), where we consider the estimates through the model of \citet{Yao_2017}. The redback sample has five parallax-based distance estimates and nine distance estimates through DM or optical light-curves. Lastly, the LMXB sample has nine parallax distances and ten distances through optical light-curves or Type I X-ray bursts \citep[see][and references therein]{ODoherty_2023}.

\begin{figure*}
    \centering
    \includegraphics[width=18cm]{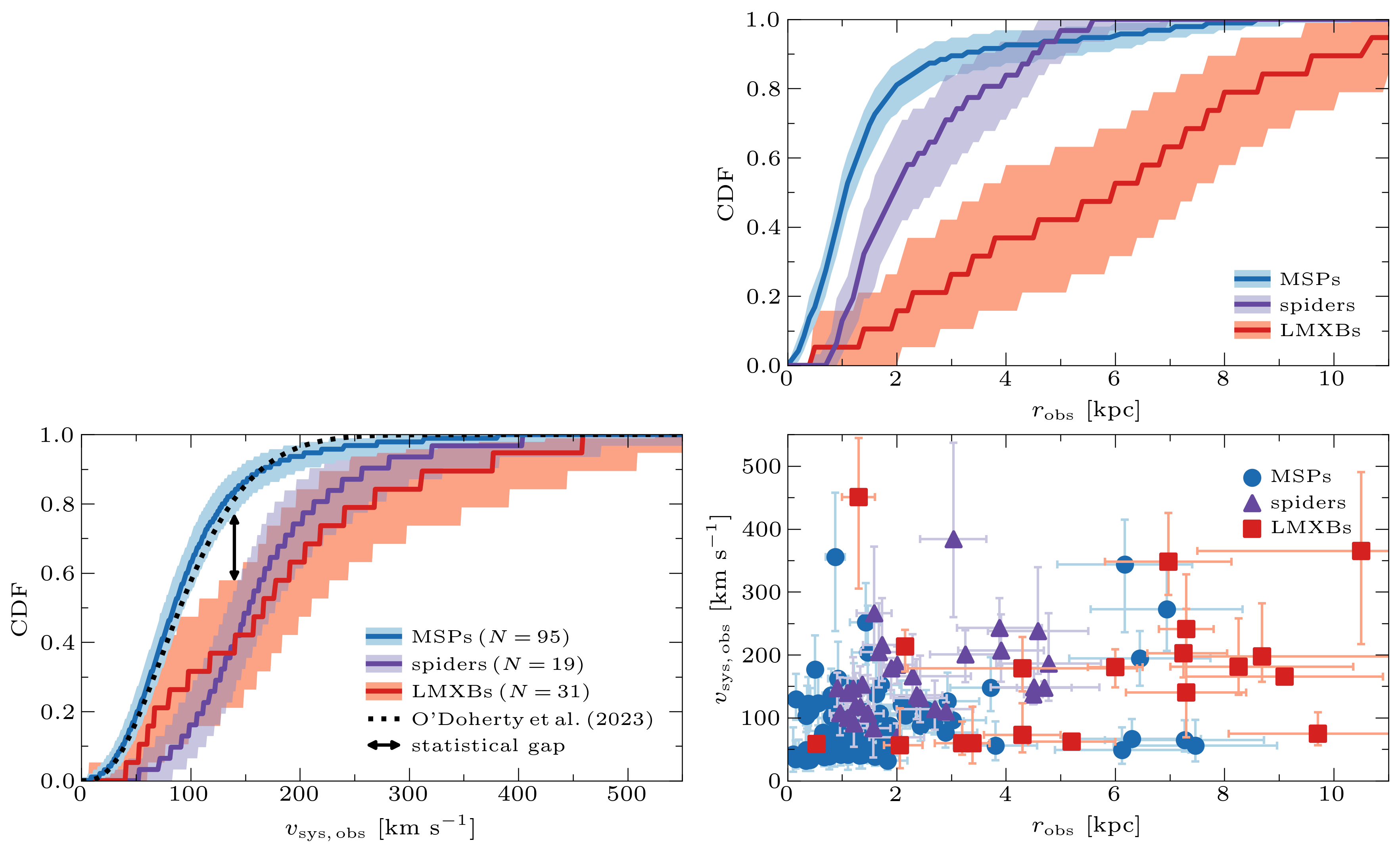}
    \caption{Observed distributions of systemic kicks (bottom left panel) and distances (top right panel) for the MSPs (blue), spiders (purple) and LMXBs (red) in the samples of \citet{ODoherty_2023}. The dark lines show the median distributions and the shaded regions correspond to the bootstrapped $90\%$ confidence intervals. The dotted line shows the fitted Beta distribution of \citet{ODoherty_2023}, which is dominated by the MSPs, and the black double-headed arrow illustrates the statistical gap between the MSPs and the other objects. The bottom right panel shows the relationship between the distances and systemic kicks for the MSPs (blue circles), spider pulsars (purple triangles), and LMXBs (red squares), where the error bars show the $68\%$ confidence intervals.}
    \label{fig1}
\end{figure*}

We use the distances and systemic kick estimates listed by \citet{ODoherty_2023} and bootstrap them in order to determine the confidence intervals of the distance distributions. In particular, we approximate each distance or kick value as a sum of two half-Gaussians with the same mean but different standard deviations, as formulated by \citet{Disberg_2023}, based on the medians and $68\%$ uncertainties. If a data set has $N$ objects, we sample (with replacement) $N$ objects from the data set, and for each object sample a distance or kick value from the corresponding asymmetric Gaussian distribution. We repeat this $10^3$ times and thus obtain $10^3$ cumulative density distributions (CDFs), allowing us to determine the median distribution and the $90\%$ confidence interval.

Figure \ref{fig1} shows the kick and distance distributions for the MSPs, spider pulsars, and LMXBs. Because the distributions for the redbacks and the black widows are very similar, we combine them into a single sample to improve population statistics.

The MSPs are closest to the Sun, with observed distances ($r_{\text{obs}}$) of ${\sim}\,1\,$kpc. The radio surveys that discover these pulsars are indeed only sensitive at relatively short distances. The spider pulsars, however, are situated at statistically significantly larger distances (${\sim}\,2\,$kpc). This may be explained by the fact that some spider pulsars are discovered through other wavelengths first before confirmation through radio observation \citep[see][and references therein]{Thongmeearkom_2024}. The LMXBs, in turn, are observed at significantly greater distances from the Sun, with the majority at $r_{\text{obs}}\gtrsim5\,$kpc, since the X-ray emission from these objects is observable at greater distances.

Figure \ref{fig1} also shows the median systemic kick distributions and their $90\%$ confidence intervals, based on the results of \citet[][see also their figure 6]{ODoherty_2023}. The figure shows that the systemic kicks of the observed LMXBs and spiders are similar, with a median kick of ${\sim}\,150\,$km\,s$^{-1}$. However, the systemic kicks of the MSPs are lower, with a median of ${\sim}\,100\,$km\,s$^{-1}$ \citep[see also][]{Ding_2025} and a noticeable gap between the confidence intervals of the MSPs and the LMXBs.

One factor that may help explain low MSP distances is the difference between distance estimates through DM and parallax, since \citet{Disberg_2025b} find that parallax distances tend to exceed DM distances. \citet{ODoherty_2023} discuss the effect of parallax distance estimates and state that using parallax distances can make the MSP systemic kicks more consistent with the black widow kicks. Although they claim that the MSP kick distributions for the DM distance sample and the parallax distance sample are similar, we note that this is not a completely fair comparison because parallax estimates are only feasible for nearby pulsars. In Appendix \ref{appA} we compare the DM and parallax distances for the $33$ MSPs in the sample where both are available, and show that parallax distances indeed lead to slightly higher systemic kick estimates than DM distances.

However, we argue that a key reason why the observed MSP kicks are significantly lower than the LMXB and spider kicks is the fact that the observed MSPs are located significantly closer to the Sun than spiders and LMXBs. Objects that receive large systemic kicks will migrate to larger distances from their birth location (i.e., the Galactic thin disc), which means that, at least as a crude approximation, one should expect to be able to see objects that have received higher kicks at larger distances. Likewise, if observations are only limited to the objects closest to the Sun, there is a \textquotedblleft kinematic bias\textquotedblright\ towards the objects that have received the lowest kicks.

As evidence for this kinematic bias, we consider the relationship between the systemic kick estimates and the distances of the observed objects. In Figure \ref{fig1} we show that the systemic kicks indeed increase for larger distances, except for a few outliers. We argue that the distributions of MSPs and LMXBs are not necessarily incompatible: the gap between the MSP and LMXB kicks seems to be caused by the different $r_{\text{obs}}$ distributions. We note that the spider pulsars, on the other hand, do seem to be concentrated at slightly higher kicks relative to the MSPs at the same distance.

\begin{figure*}
    \centering
    \includegraphics[width=18cm]{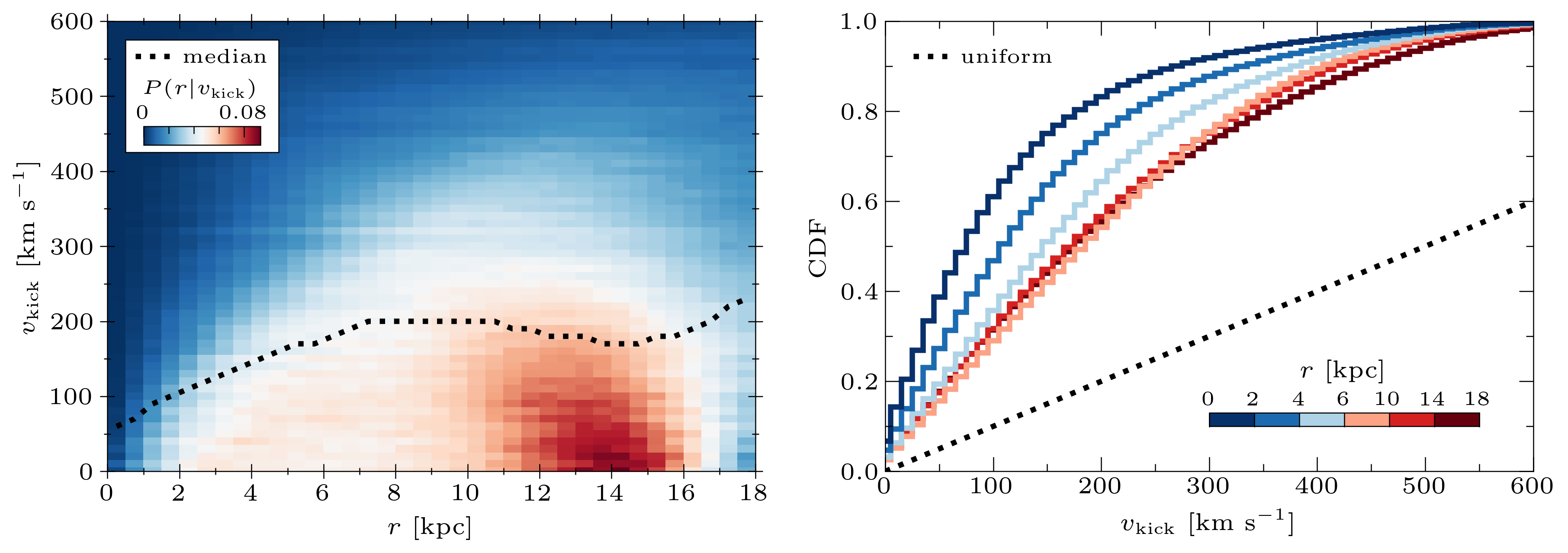}
    \caption{Distances of the kicked objects in the simulation of \citet{Disberg_2025a} for $40\,$Myr$\,{<}\,t\,{<}\,1\,$Gyr (evaluated with timesteps of $1\,$Myr). The left panel shows, for each simulation containing $10^3$ objects with a kick magnitude $v_{\text{kick}}$, the distribution of distances to the Sun ($r$) in a 2D histogram with distance bins of $0.5\,$kpc and velocity bins of $10\,$km\,s$^{-1}$. The rows are normalized so that the distribution corresponds to $P(r\giv v_{\text{kick}})$. The dotted line shows the median kick velocity in each distance bin. The right panel shows kick distributions integrated over certain distance ranges (colors) through Equation (\ref{eq1}) for a uniform kick distribution between $0$ and $1000\,$km\,s$^{-1}$, along with that uniform distribution itself (dotted line), which is the kick distribution observed at $t=0$ for any distance. We note that (1) the simulation considers kick velocities up to $1000\,$km\,s$^{-1}$, even though the figure shows kicks up to $600\,$km\,s$^{-1}$, and (2) $v_{\text{kick}}$ equals the natal kick for single objects but the systemic kick for binaries.}
    \label{fig2}
\end{figure*}

\section{Simulation}
\label{sec3}
\noindent In order to illustrate this distance-dependent kinematic bias, we simulate the kick-dependent spatial distribution of kicked objects. This allows us to estimate the expected kick distribution for a population of systems following an observed distance distribution, given an intrinsic kick distribution. Comparison between this intrinsic distribution and the kick distribution expected to be observed (given a distance distribution) describes the kinematic bias. Through such a comparison, we aim to investigate whether a single intrinsic kick distribution can explain the observed kick distributions shown in Figure \ref{fig1} by applying a distance-dependent kinematic bias.

This simulation enables us to estimate the distance distribution for a certain kick value, $P(r\giv v_{\text{kick}})$, which is related to the distance-dependent kick distribution through $P(v_{\text{kick}}\giv r)=P(v_{\text{kick}})P(r\giv v_{\text{kick}})/P(r)$. Here, $P(v_{\text{kick}})$ is the intrinsic kick distribution and $P(v_{\text{kick}}\giv r)$ is the kick distribution of objects at a distance $r$. The expected observable kick distribution of objects found at $r_{\min}\,{<}\,r\,{<}\,r_{\max}$ is given by
\begin{equation}
    \label{eq1}
    P(v_{\text{kick}}\giv r_{\min},r_{\max})= P(v_{\text{kick}}) \int_{r_{\min}}^{r_{\max}}\dfrac{P(r\giv v_{\text{kick}})}{P(r)}dr,
\end{equation}
where $v_{\text{kick}}$ corresponds to the natal kick for isolated objects but the systemic kick for binaries. We note that young objects are still close to their birth location independent of their kick magnitude, meaning $P(r\giv v_{\text{kick}})\approx P(r)$. However, for older objects (such as the ones in the samples shown in Figure \ref{fig1}) the observed locations differ from the birth locations, depending on the kick magnitude. The method of estimating the observed kick distribution solely based on distance is limited by the fact that it does not take into account the sky location, which could provide additional information, but we use this method to illustrate the effect of the kinematic bias.

In order to estimate $P(r\giv v_{\text{kick}})/P(r)$, we employ the simulation of \citet{Disberg_2025a}, who seed $10^3$ point masses in the thin disc in a circular Galactic orbit and add an isotropic kick velocity. Then, they use \lstinline{GALPY}\footnote{\href{http://github.com/jobovy/galpy}{http://github.com/jobovy/galpy}} \citep{Bovy_2015} to integrate the Galactic trajectories of the kicked objects through the Galactic potential \citep[using the potential of][]{McMillan_2017}. They integrate these trajectories for $1\,$Gyr, and repeat the simulation for different kick velocities, ranging from $0$ to $1000\,$km\,s$^{-1}$ with steps of $10\,$km\,s$^{-1}$ (effectively uniformly distributed). Moreover, they take the current position and velocity of the Sun \citep[see][and references therein]{Disberg_2024a} and integrate its trajectory through the Galaxy along with the kicked objects, based on which the distance between the Sun and the objects can be determined at each timestep. We consider the distances of the simulated objects at $t>40\,$Myr, since after this point the system has reached dynamical equilibrium \citep{Disberg_2024a,Disberg_2024b} and LMXBs, spiders, and MSPs are all expected to be dynamically old (i.e., ${>}\,40\,$Myr).

A key assumption in this simulation is the birth locations of the kicked objects. \citet{Disberg_2024b} find that the birth locations do not affect the eccentricity of the Galactic trajectories of observed objects, but we are interested in how the kick magnitude affects the distances to the objects, which is significantly more sensitive to the initial conditions. We choose to follow the birth locations used by \citet{Disberg_2025a}, who seed their simulated objects in the thin disc. For the initial locations they use the Gaussian annulus distribution of \citet{Faucher_2006}, who estimate the birth locations of NSs based on the pulsar data of \citet{Yusifov_2004}. In this distribution, the points are seeded at $z=0$ in an annulus formulated as \citep[see also, e.g.,][]{Sartore_2010}
\begin{equation}
    \label{eq2}
    P(R)\propto\exp\left(-\dfrac{(R-7.04\,\text{kpc})^2}{2(1.83\,\text{kpc})^2}\right),
\end{equation}
where $R$ is the galactocentric radius. Indeed, massive stars are formed in the spiral arms of the Galaxy at Galactic height $|z|\,\lesssim50\,$pc \citep[e.g.,][]{Bronfman_2000,Urquhart_2011,Urquhart_2014}, and it is this population that produces NSs. Moreover, \citet{ODoherty_2023} assume that the objects in their samples are formed in the disc, since they consider peculiar velocities at disc crossings as potential kick velocities. In Appendix \ref{appB} we repeat the simulation but for a different birth location distribution, which is shifted towards the center of the Galaxy, and show that this reduces the effect of the kinematic bias particularly at $r\lesssim6\,$kpc.

Figure \ref{fig2} shows the relationship between the distances to the simulated objects and kick velocity. The left panel shows the number of objects at a certain distance $r$ from the Sun, depending on kick magnitude, normalized such that it corresponds to $P(r\giv v_{\text{kick}})$. The figure shows that, for a uniform intrinsic kick distribution, the median observed kick increases from ${\sim}\,50\,$km\,s$^{-1}$ at ${\sim}\,0\,$kpc to ${\sim}\,200\,$km\,s$^{-1}$ at ${\gtrsim}\,7\,$kpc, due to the distance-dependent kinematic bias. Most objects are situated at relatively large distances because the surface of a sphere with radius $r$ scales with $r^2$. However, we are not interested in the distance distribution but in the kick distribution of objects found at a certain distance, corresponding to $P( v_{\text{kick}}\giv r)$ as in Equation (\ref{eq1}). This forms a prior expectation on what kicks one would observe when looking at objects at that distance (without any knowledge about the intrinsic kick distribution). 

In the right panel of Figure \ref{fig2} we show kick distributions integrated over several distance ranges through Equation (\ref{eq1}), assuming a uniformly distributed intrinsic kick distribution. Here, we estimate $P(r\giv v_{\text{kick}})$ through our simulation, which is displayed in the left panel of Figure \ref{fig2}. Similarly, $P(r)$ scales with the total value of a column. \citet{Disberg_2025a} find that after $40\,$Myr the objects observed within $2\,$kpc of the Sun are biased towards low kicks, but Figure \ref{fig2} shows that this kinematic bias is indeed distance dependent. For a population of objects with a certain kick distribution, the nearby objects have experienced lower kicks than the ones further away. High kicks are likely to eject objects from the Galaxy, where the escape velocity near the Sun is ${\sim}\,530\,$km\,s$^{-1}$ \citep{Deason_2019}, so they are not observed in the Galactic population after $40\,$Myr.

\begin{figure*}
    \centering
    \includegraphics[width=18cm]{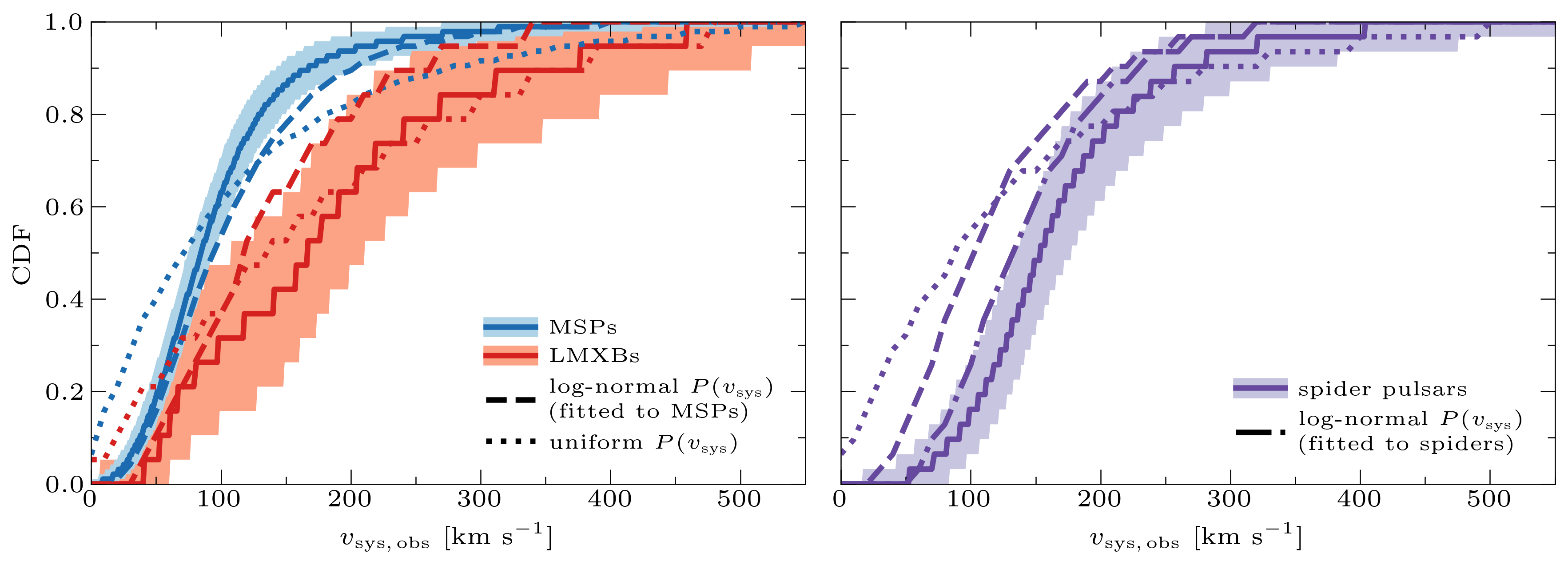}
    \caption{Observed systemic kick distributions of \citet{ODoherty_2023}, for MSPs (blue), LMXBs (red), and spider pulsars (purple). The solid lines show the median distributions and the shaded regions correspond to the $90\%$ confidence intervals. The dashed lines are determined through Equations (\ref{eq3}) and (\ref{eq4}) and a lognormal $P(v_{\text{sys}})$ distribution fitted to the lower edge of the MSP confidence interval. The dash-dotted line, in turn, uses a $P(v_{\text{sys}})$ distribution fitted to the upper edge of the spider pulsars confidence interval. The dotted line shows the kinematic bias for a uniformly distributed $P(v_{\text{sys}})$.}
    \label{fig3}
\end{figure*}

\section{Reconciliation}
\label{sec4}
\noindent In order to compare the simulation to the observed kick distributions of \citet{ODoherty_2023}, we estimate what kicks one should expect to observe given (1) the distances of the observed objects (as shown in Figure \ref{fig1}), (2) the distance-dependent kinematic bias (shown in Figure \ref{fig2}), and (3) an assumed intrinsic systemic kick distribution. In particular, we choose a model $\mathcal{M}$ that describes the intrinsic kick distribution $P(v_{\text{kick}}\giv\mathcal{M})$. Then, the distribution of kicks for objects located at a distance $r$ is given by
\begin{dmath}
    \label{eq3}
    P(v_{\text{kick}}\giv r,\mathcal{M})=\dfrac{P(v_{\text{kick}}\giv\mathcal{M})P(r\giv v_{\text{kick}})}{\int P(v_{\text{kick}}\giv\mathcal{M})P(r\giv v_{\text{kick}})dv_{\text{kick}}}.
\end{dmath}
In practice, we do not know the distance for a given object perfectly, but have a probability distribution for the distance $P(r \giv \{r_{\text{obs}}\})$ based on distance observations $\{r_{\text{obs}}\}$ which we describe through their median and uncertainty (as described in Section \ref{sec2}). We can use this, together with Equation (\ref{eq3}), to predict the expected kick velocity posterior for an object given its distance measurements and the intrinsic kick model, which we formulate as
\begin{dmath}
    \label{eq4}
    P(v_{\text{kick}}\giv\{r_{\text{obs}}\},\mathcal{M})=\int P(v_{\text{kick}}\giv r,\mathcal{M})P(r\giv\{r_{\text{obs}}\})dr.
\end{dmath}
The value of $P(r\giv v_{\text{kick}})$ is given by our simulation and shown in the left panel of Figure \ref{fig2}. Assuming an intrinsic systemic kick model $\mathcal{M}$, we determine the expected kick distribution for each object, corrected for the kinematic bias through Equations (\ref{eq3}) and (\ref{eq4}). We repeat this $10^3$ times, resulting in $10^3$ distributions, and consider the median distributions for each category.

In an attempt to bridge the gap between the systemic kick estimates for MSPs and the ones for spiders and LMXBs, we fit the median distribution for MSPs, determined through Equations (\ref{eq3}) and ($\ref{eq4}$), to the lower edge of the $90\%$ interval of the observed MSP kicks, by choosing an intrinsic systemic kick distribution model $P(v_{\text{sys}})$. For the shape of this systemic kick distribution, we follow \citet{Disberg_2025b} and use a lognormal model, formulated as
\begin{equation}
    \label{eq5}
    P(v_{\text{sys}}\giv\mu,\sigma)=\dfrac{1}{v_{\text{sys}}\,\sigma\sqrt{2\pi}}\exp\left(-\dfrac{(\ln v_{\text{sys}}-\mu)^2}{2\sigma^2}\right),
\end{equation}
where $v_{\text{sys}}$ given in km\,s$^{-1}$, and $\mu$ and $\sigma$ describe the kick model $\mathcal{M}$ in Equations (\ref{eq3}) and (\ref{eq4}). Fitting this to the lower edge of the MSP interval yields $\mu=5.1$ and $\sigma=0.7$, resulting in a median kick velocity of $164\,$km\,s$^{-1}$. In Appendix \ref{appC} we justify our choice of a lognormal distribution by showing that if we fit an alternative model of $P(v_{\text{kick}}\giv\mathcal{M})$, a histogram with bins of $50\,$km\,s$^{-1}$, to the observed kicks, the result is similar to the lognormal fit.

\begin{figure*}
    \centering
    \includegraphics[width=18cm]{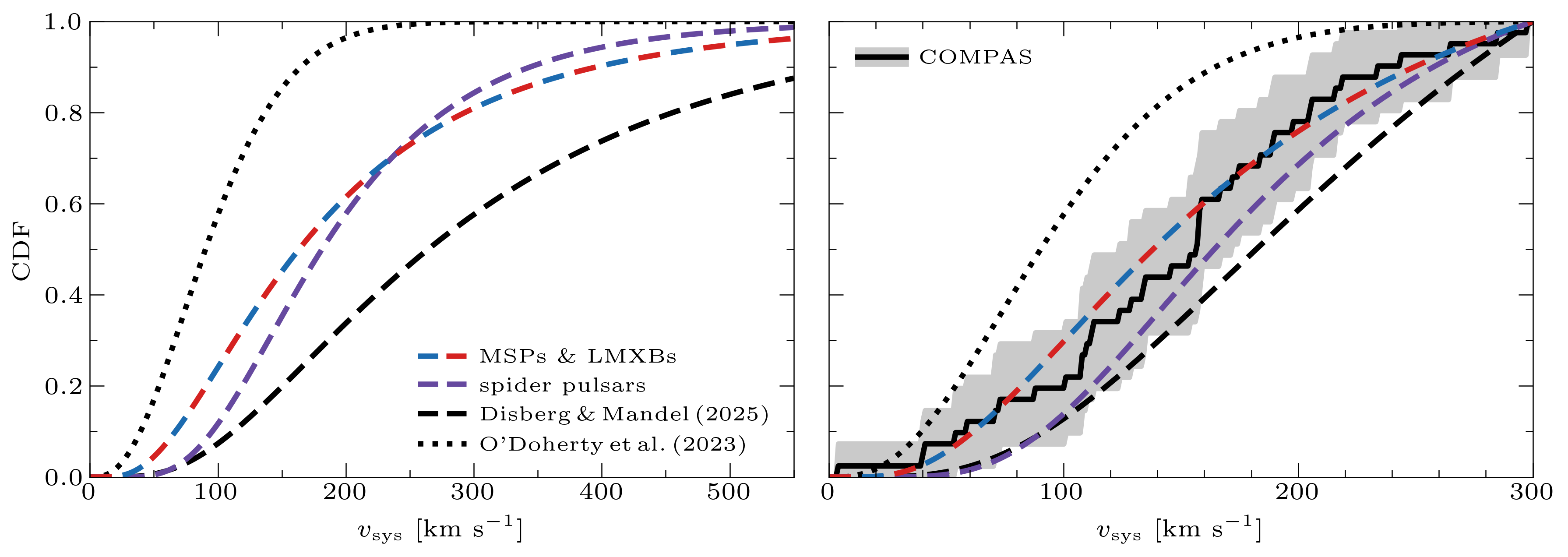}
    \caption{Systemic kick distributions fitted to the observations while accounting for the kinematic bias. The left panel shows these lognormal distributions, for the MSPs and LMXBs ($\mu=5.1$ and $\sigma=0.7$, blue and red dashed line), as well as for the spider pulsars ($\mu=5.2$ and $\sigma=0.5$, purple dashed line). The dashed line shows the natal NS kick distribution of \citet{Disberg_2025b}, fitted to isolated pulsar velocities, while the dotted line shows the fit of \citet{ODoherty_2023} to their observed systemic kick estimates of MSPs, spiders, and LMXBs (as shown in Figure \ref{fig3}). The right panel shows the same distributions, but normalized between $0$ and $300\,$km\,s$^{-1}$. The shaded region is the $95\%$ bootstrapped interval of COMPAS binaries with $P_{\text{orb}}\leq10\,$d and $M_{c}\leq1\,M_{\odot}$ (as discussed in the text).}
    \label{fig4}
\end{figure*}

The dashed lines in Figure \ref{fig3} show these median observed kick distributions, based on their distances and our lognormal model. The MSP prediction indeed traces the lower edge of the MSP confidence interval, whereas the LMXB prediction, which combines the same intrinsic kick distribution with the LMXB distances, traces the upper edge of the LMXB interval. In other words, in this model the kinematic bias can close the gap between the observed MSP and LMXB kicks, merely based on the fact that the LMXBs have larger distances.

However, if we apply this lognormal kick distribution to the spider pulsars, we find that the prediction underestimates the observed kicks. Even though the observed kicks of the spider pulsars are similar to the LMXB kicks, they are difficult to reconcile using our model because the spiders are located at significantly lower distances (Figure \ref{fig1}). Instead, we fit a lognormal kick distribution to the upper edge of the spider interval, finding $\mu=5.2$ and $\sigma=0.5$, with a median kick of $181\,$km\,s$^{-1}$.

\section{Comparison}
\label{sec5}
\noindent We find that a single intrinsic systemic kick distribution can reconcile the MSP and LMXB systemic kick estimates. The spider kicks are best fit by a slightly different intrinsic distribution, but it is unclear whether this difference is statistically significant, considering the systematic uncertainty in the distance estimates for these objects and the limitations in our model by only considering $r_{\text{obs}}$, for instance. Alternatively, one could of course choose a distribution to reconcile the LMXBs and spiders and contrast this with the MSPs. In Figure \ref{fig4}, we show the systemic kick distributions, together with the distribution that \citet{ODoherty_2023} fitted to the observed systemic kicks (as shown in Figure \ref{fig3}) as well as the natal kick distribution of \citet{Disberg_2025b}. Our intrinsic systemic kick distributions are higher than the observations of \citet{ODoherty_2023}, because they are corrected for kinematic bias through Equation (\ref{eq3}), and they are lower than the natal kicks of \citet{Disberg_2025b}, because these objects are binaries that survive the natal kick.

We are interested in estimating whether the systemic kick distributions we show in Figure \ref{fig4} are consistent with the systemic kicks we would expect based on binary evolution \citep[e.g.,][]{Kalogera_1998}. To this end, we employ \lstinline{COMPAS} rapid binary population synthesis v03.22.01 \citep{Stevenson_2017,Vigna_2018,COMPAS_2022a,COMPAS_2025}. We use default settings, but change the $v_{\text{NS}}$ and $\sigma_{\text{kick}}$ parameters in the kick prescription of \citet{Mandel_2020} to $630\,$km\,s$^{-1}$ and $0.45$, respectively \citep[cf.][]{Kapil_2023}, which were fitted to the results of \citet{Disberg_2025b}. We simulate a million binaries, and select the ones that (1) form an NS, (2) have a companion that is not a compact object, with a mass of $M_{c}\leq1\,M_{\odot}$, (3) remain bound after the natal kick, (4) have a post-supernova orbital period of $P_{\text{orb}}\leq10\,$d. This selection yields $41$ binaries that we consider as potential progenitors of LMXBs, illustrating that formation through isolated binary evolution is rare.

We bootstrap the systemic kicks of these binaries, as explained in Section \ref{sec2}, and in the right panel of Figure \ref{fig4} we show the corresponding confidence interval. The observations have a high velocity tail due to projection effects which are not present in the \lstinline{COMPAS} results, because of which we normalize these distributions between $0$ and $300\,$km\,s$^{-1}$ to investigate the distribution around their peaks. As the figure shows, the systemic kick distributions inferred from observations after accounting for kinematic bias are consistent with the \lstinline{COMPAS} results. On the other hand, the isolated pulsar velocity distribution of \citet{Disberg_2025b} has larger velocities because binary survival and the inertial anchor of a companion leads to lower binary systemic kicks than isolated neutron star natal kicks. Meanwhile, the total distribution inferred by \citet{ODoherty_2023} without accounting for the kinematic bias lies above the confidence interval.

In this work, we used the current \lstinline{COMPAS} default, other than the modifications to the \citet{Mandel_2020} kick scaling as described above \citep[as well as the core-mass prescription of][]{Brcek_2025}. Variations in uncertain physical assumptions, such as those governing mass transfer and the common-envelope phase, do impact the kick distributions; however, a partial exploration of the large parameter space suggests that the key results reported here are not affected.

Figure \ref{fig4} shows that the difference between the systemic kick distribution for MSPs and LMXBs and the distribution for spider pulsars is relatively small. We argue that systematic uncertainty in distance estimates is likely sufficient to reconcile this difference (see, e.g., Appendix \ref{appA}). Moreover, we find that the \lstinline{COMPAS} results are relatively sensitive to the companion mass. In Appendix \ref{appD} we show that lower companion masses indeed yield higher systemic kicks, likely because of the fact that mass transfer onto lower mass stars yields tighter orbits with higher pre-supernova orbital velocities \citep[see, e.g.,][]{Mandel_2026}. We hypothesize that it may be possible that spider pulsars have an observational bias towards lower companion masses, since these are easier to ablate and might create stronger optical signals, which may affect their observed systemic kick distribution.

\section{Conclusion}
\label{sec6}
\noindent Based on our analysis of the MSP, spider pulsar, and LMXB systemic kicks as determined by \citet{ODoherty_2023}, we conclude the following:
\begin{itemize}
    \item Objects that receive higher (systemic) kicks reach greater distances from the Sun. Therefore, for a given kick distribution, dynamically old objects (i.e., older than $40\,$Myr; \citealt{Disberg_2024a,Disberg_2024b}) found nearby likely received lower kicks.
    \item We model this kinematic bias through the simulation of \citet{Disberg_2025a}, effectively assuming that the birth locations of the simulated objects follow Equation (\ref{eq2}) and are thus born in the Galactic disc \citep{Faucher_2006}. Through this model, we find a systemic kick distribution that can bridge the gap between the observed kicks of MSPs and LMXBs, based on the fact that the MSPs are located at significantly shorter distances than the LMXBs. This distribution is described by a lognormal model with $\mu=5.1$ and $\sigma=0.7$, resulting in a median kick velocity of $164\,$km\,s$^{-1}$.
    \item This kick distribution predicts slightly lower systemic kicks for spider pulsars than observed, possibly due to uncertainty in distance measurements or a hypothetical observational bias towards lower companion masses.
    \item Our systemic kick distributions for MSPs, spiders, and LMXBs are consistent with simulated \lstinline{COMPAS} binaries with $P_{\text{orb}}\leq10\,$d and $M_{c}\leq1\,M_{\odot}$, where the natal kicks are determined through the model of \citet{Mandel_2020} calibrated with the results of \citet{Disberg_2025b}.
\end{itemize}
In general, we conclude that taking into account the different distances of the observed MSP, spider, and LMXB populations can contribute to reconciling their systemic kicks: this kinematic bias can be sufficient to bridge the gap between the MSP and LMXB confidence intervals. Based on this, as well as systematic uncertainties in the distance estimates and possible observational biases, we argue that even though \citet{ODoherty_2023} find significantly lower kicks for MSPs than for spiders and LMXBs, this does not prove that they come from different systemic kick distributions. The results of \citet{ODoherty_2023} are therefore not in tension with a common origin of MSPs, spider pulsars, and LMXBs.

\begin{acknowledgments}
\noindent We thank the referee for helpful comments, and Tyrone O'Doherty and Alexey Bobrick for useful discussions.\ P.D.\ and I.M.\ acknowledge support from the Australian Research Council (ARC) Centre of Excellence for Gravitational-Wave Discovery (OzGrav) through project number CE230100016.
\end{acknowledgments}
\noindent\software{\lstinline{ASTROPY} \citep{Astropy_2013,Astropy_2018,Astropy_2022}, \lstinline{COMPAS} \citep{COMPAS_2022a,COMPAS_2022b,COMPAS_2025}, \lstinline{GALPY} \citep{Bovy_2015}, \lstinline{MATPLOTLIB} \citep{Hunter_2007}, \lstinline{NUMPY} \citep{Harris_2020}, \lstinline{SCIPY} \citep{Virtanen_2020}.}

\clearpage
\appendix
\section{Parallax Distances}
\label{appA}
\renewcommand{\thefigure}{A}
\noindent The distance estimates of the MSPs in the sample of \citet{ODoherty_2023} are based on DM, as discussed in Section \ref{sec2}. However, \citet{Disberg_2025b} find that parallax distances, which may be more reliable \citep[e.g.,][]{Deller_2009,Verbunt_Cator_2017,Deller_2019}, tend to exceed DM distances for the sample of isolated pulsars with both distance measurements available. The observed kick velocity is proportional to the distance for a fixed (and typically well-measured) proper motion. Therefore, we consider the $33$ MSPs from the sample that have parallax estimates and compare their systemic kick estimates from \citet{ODoherty_2023} for DM distances and parallax distances. Similarly to Figure \ref{fig3}, we bootstrap the kick estimates to determine the $90\%$ confidence interval. In Figure \ref{figA} we show these distributions, together with the LMXB distribution. The MSP kicks based on parallax distances indeed exceed the ones based on DM distances, with a difference in median distribution of ${\sim}\,10{-}20\,$km\,s$^{-1}$. Although this difference is relatively small, the figure shows that it can play a significant role in bridging the gap between the MSP and LMXB confidence regions. We note, however, that part of the reason why the gap is smaller for the parallax MSPs (compared to Figure \ref{fig1}) is because of the fact that the smaller sample size ($33$ compared to $95$) leads to a larger uncertainty region.

\begin{figure}[ht]
    \centering
    \resizebox{\hsize}{!}{\includegraphics{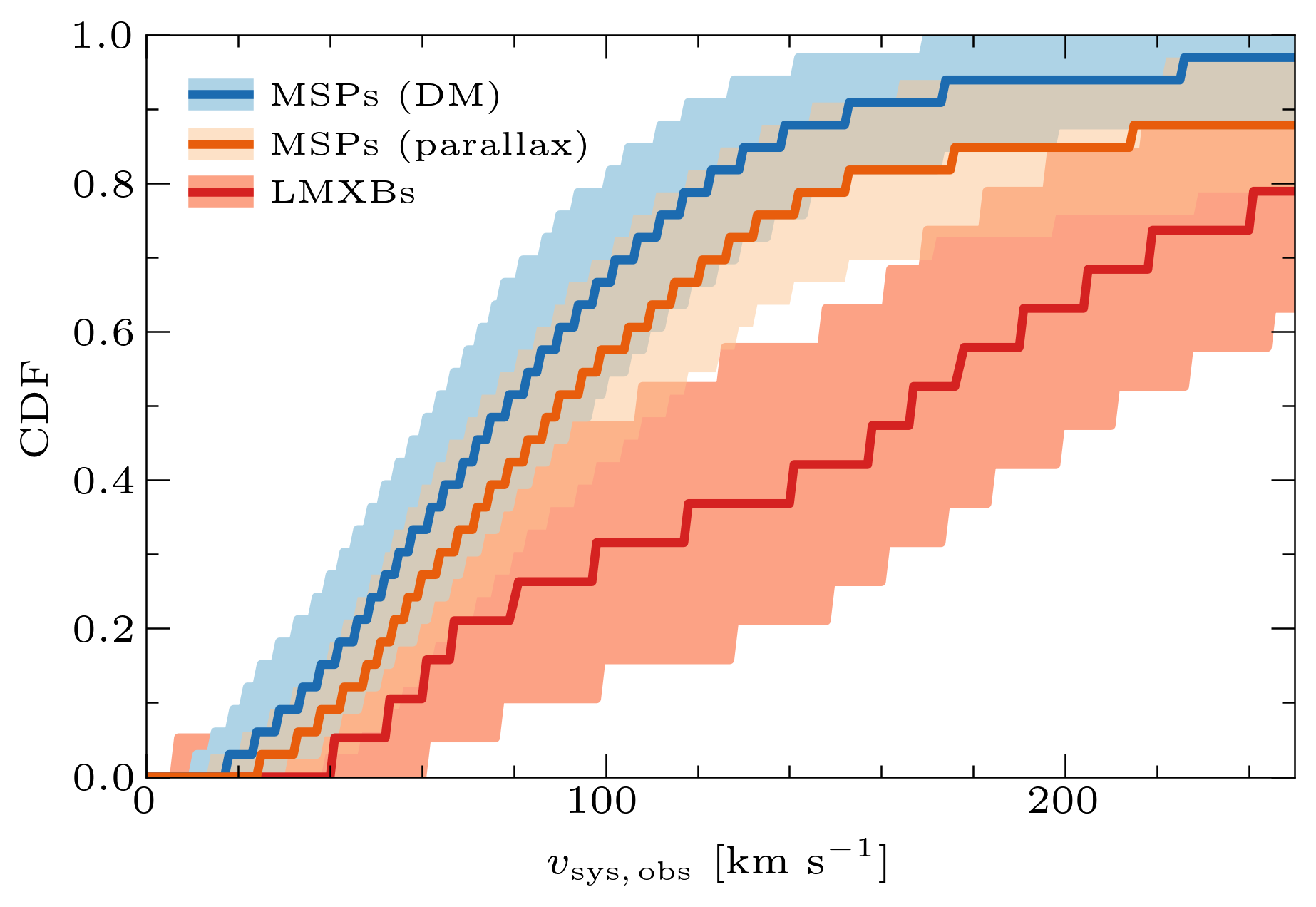}}
    \caption{Systemic kick distributions from \citet{ODoherty_2023} for $33$ MSPs with parallax estimates, based on DM distances (blue) and parallax distances (orange), as well as the distribution of LMXBs (red) similarly to Figure \ref{fig3}. The dark lines show the median distributions and the shaded regions show the bootstrapped $90\%$ confidence intervals.}
    \label{figA}
\end{figure}

\newpage
\section{Birth Locations}
\label{appB}
\noindent The simulation discussed in Section \ref{sec3} seeds point masses in a Gaussian annulus \citep{Faucher_2006} described by Equation (\ref{eq2}), following massive star formation in the Galactic thin disc \citep[e.g.,][]{Bronfman_2000,Urquhart_2011,Urquhart_2014}. In order to investigate the effect of this assumption, we repeat the simulation but with birth locations more concentrated in the Galactic bulge. We describe these birth locations using the Milky Way model of \citet{Chrimes_2021}. In particular, we consider the Galaxy components consisting of a triaxial boxy Gaussian bulge \citep{Grady_2020} and an exponential disc \citep{Bland_2016}. We initially weigh these components following the luminosities of \citet{Flynn_2006}, but afterwards reweigh the seeded objects such that effectively $10^3$ are born in the Galactic bulge (which we define as $R\leq3\,$kpc) and $10^4$ are born in the disc ($R>3\,$kpc), following the LMXB population synthesis of \citet{VanHaaften_2015}. Figure \ref{figB} shows the results for the simulation with these new birth locations, similar to the results shown in Figure \ref{fig2} but for kicks up to $600\,$km\,s$^{-1}$. The distance dependence of the kinematic bias is smaller in this simulation, particularly for $r\leq6\,$kpc. 

\renewcommand{\thefigure}{B}
\begin{figure}[ht]
    \centering
    \resizebox{\hsize}{!}{\includegraphics{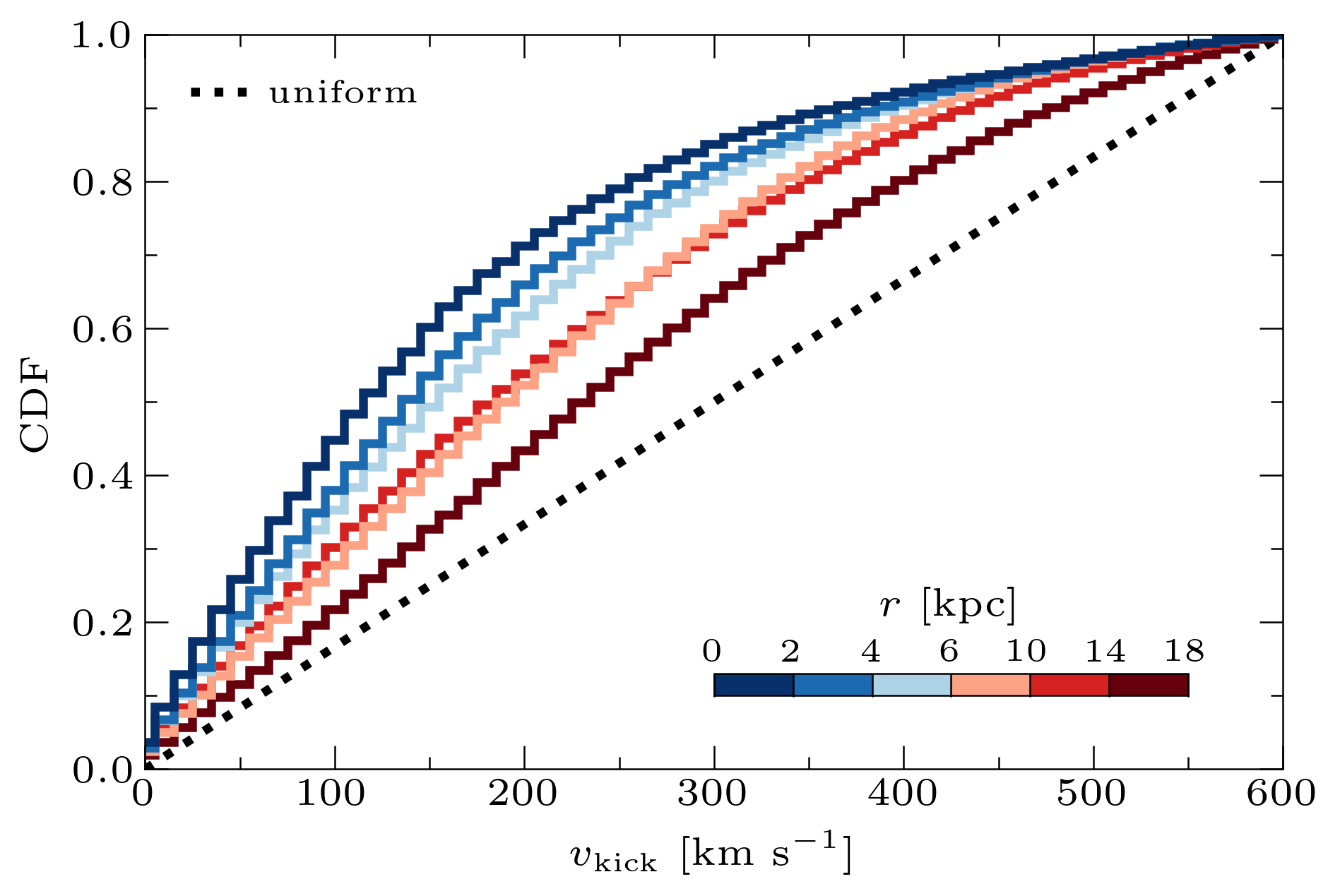}}
    \caption{Kick velocity distribution of simulated objects older than $40\,$Myr for several distance ranges (colors), similar to Figure \ref{fig2} but through a simulation using birth locations as described in the text. The dotted line shows a uniform distribution.}
    \label{figB}
\end{figure}
\renewcommand{\thefigure}{D}
\begin{figure*}
    \centering
    \includegraphics[width=18cm]{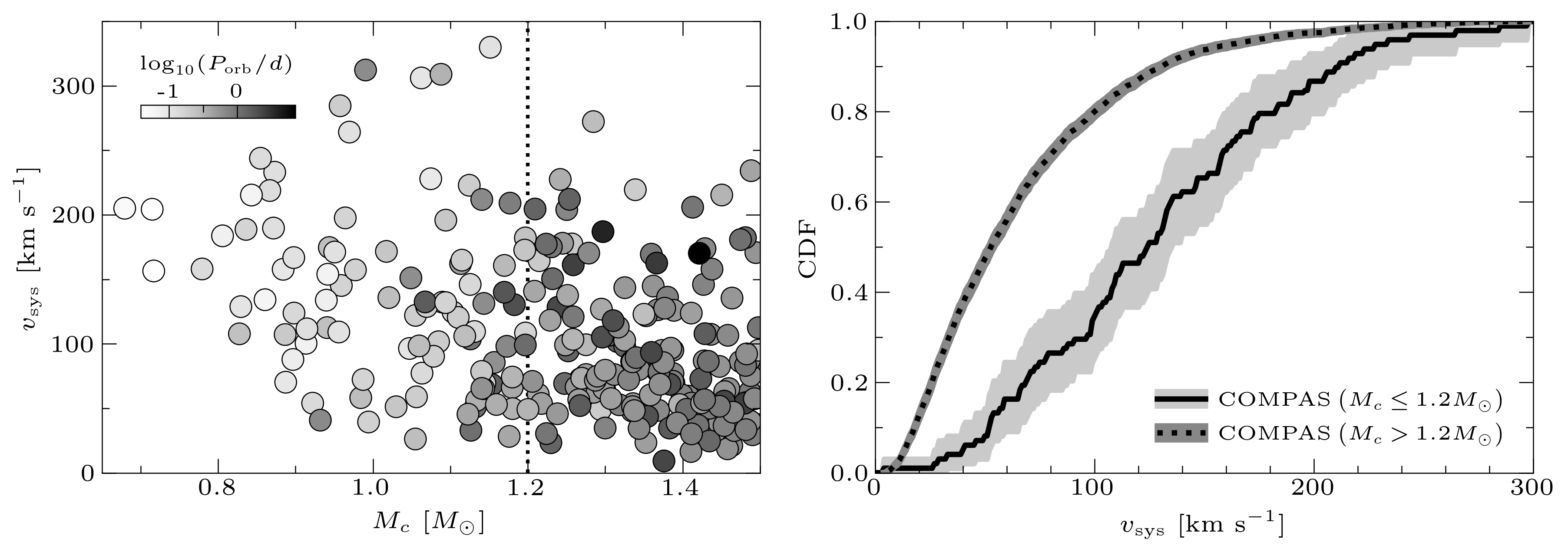}
    \caption{Relationship between the systemic kick ($v_{\text{sys}}$) and the companion mass ($M_{c}$) of the \lstinline{COMPAS} binaries described in Section \ref{sec5}. The left panel shows this relationship, where the intensity denotes the pre-supernova orbital period. We note that these binaries are selected such that their post-supernova orbital period is less than $10$ days. The right panel shows the bootstrapped confidence intervals, similarly to Figure \ref{fig4}, both for $M_{c}\leq1.2\,M_{\odot}$ (light grey region) as well as $M_{c}>1.2\,M_{\odot}$ (dark grey region), with medians described by the solid line and dotted line, respectively.}
    \label{figD}
\end{figure*}
\newpage
\section{Lognormal Model}
\label{appC}
\noindent In Section \ref{sec4} we fit a lognormal distribution to the MSP data of \citet{ODoherty_2023}, through Equations (\ref{eq3}) and (\ref{eq4}), and find that a systemic kick distribution with $\mu=5.1$ and $\sigma=0.7$ can reconcile the observed kicks for MSPs and LMXBs. Even though Figure \ref{fig3} shows that this lognormal distribution is sufficient to bridge the gap between MSPs and LMXBs, we justify our choice for a lognormal model by reiterating this fit but with a histogram with bins of $50\,$km$\,$s$^{-1}$. This results in $12$ bins between $0$ and $600\,$km$\,$s$^{-1}$, which can be used as free parameters. Figure \ref{figC} shows that the best-fitting histogram indeed follows the same shape as our lognormal fit.
\renewcommand{\thefigure}{C}
\begin{figure}[ht]
    \centering
    \resizebox{\hsize}{!}{\includegraphics{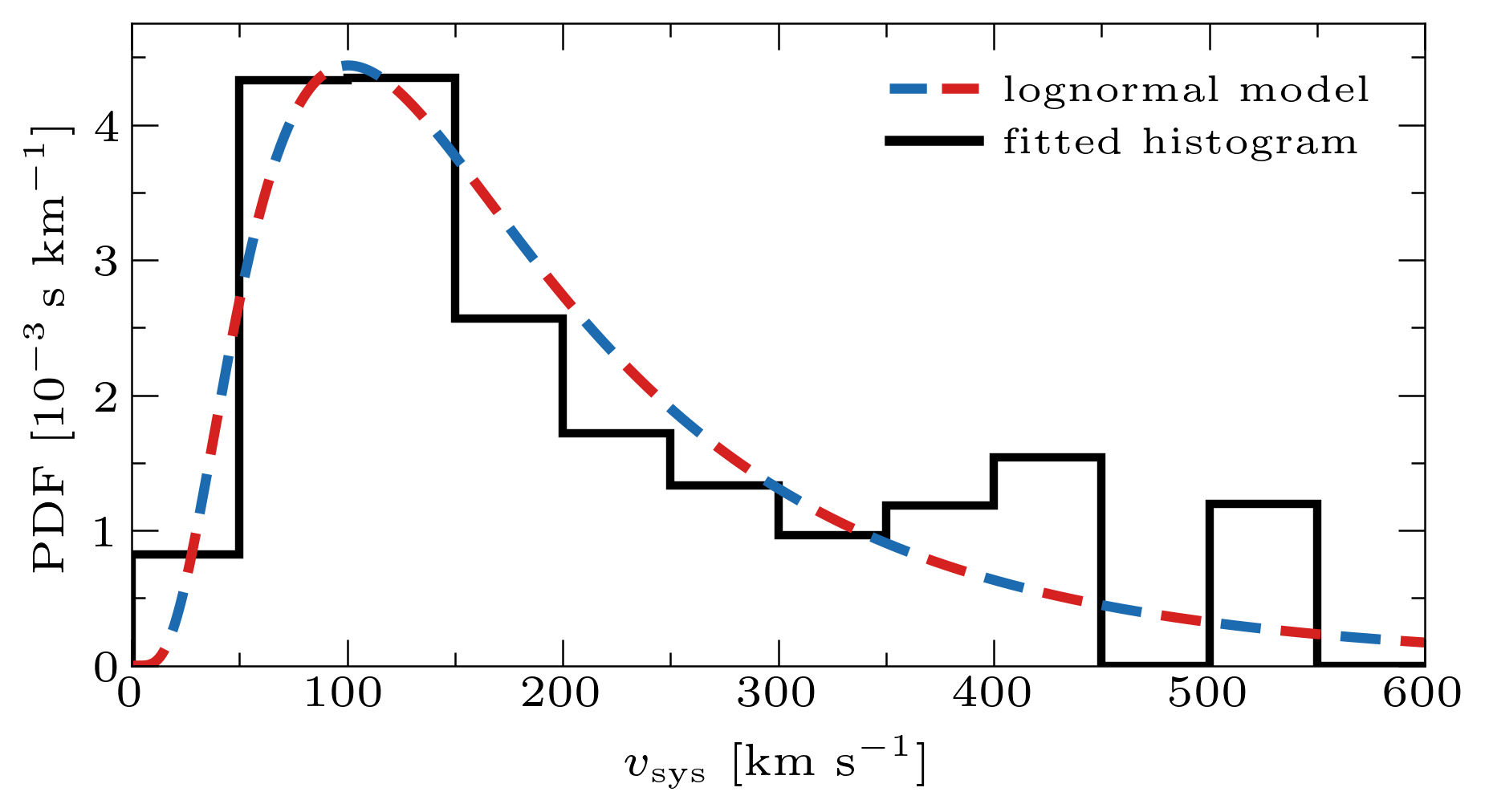}}
    \caption{Histogram fitted to the lower edge of the MSP kick confidence interval through Equations (\ref{eq3}) and (\ref{eq4}), similarly to Figure \ref{fig3}. The blue and red dashed line shows the lognormal fit.}
    \label{figC}
\end{figure}
\newpage
\section{Sensitivity to Companion Mass}
\label{appD}
\noindent In Section \ref{sec5} we show that our systemic kick distributions for MSPs, spider pulsars, and LMXBs are consistent with binaries simulated through \lstinline{COMPAS} \citep{COMPAS_2022a,COMPAS_2022b,COMPAS_2025}. We note, however, that the systemic kick distribution of these simulated binaries is relatively sensitive to the companion mass. In Figure \ref{figD}, we show the relationship between systemic kick and companion mass. The systemic kick increases for lower companion mass, likely because these binaries have shorter pre-supernova orbital periods because mass transfer from the progenitor of the neutron star onto a smaller companion leaves the binary on a tighter orbit. A shorter pre-supernova orbital period means that the binary (1) can experience higher natal kicks without unbinding \citep[e.g.,][]{Mandel_2026}, and (2) has a higher orbital velocity and therefore a higher Blaauw kick \citep{Blaauw_1961}. To illustrate this dependence on companion mass, we repeat the bootstrapping of these systemic kicks but for $M_{c}\leq1.2\,M_{\odot}$ and $>1.2\,M_{\odot}$. As the right panel in Figure \ref{figD} shows, the binaries with the lowest companion masses indeed have significantly higher systemic kicks.

\clearpage
\bibliography{references}{}
\bibliographystyle{aas_v7}

\end{document}